\documentclass[authoryear,12pt]{elsarticle}
\usepackage{graphicx}
\usepackage{lineno}
\usepackage{float}
\def\lsim{\ ^{<}\!\!\!\!_{\sim}\>}
\def\gsim{\ ^{>}\!\!\!\!_{\sim}\>}
 
\begin{document}


\centerline{\Large\bf The depletion of the asteroid belt and the}

\centerline{\Large\bf impact history of the Earth}

\vspace{2cm}

\centerline{Julio A. Fern\'andez$^{(*)}$}

\noindent Departamento de Astronom\'ia, Facultad de Ciencias, Universidad de la Rep\'ublica, Igu\'a 4225, 14000 Montevideo, Uruguay

\vspace{4cm}

\centerline{September 16, 2025}

\vspace{4cm}

\centerline{ICARUS, in press}

\vspace{4cm}

\noindent $^{(*)}$ Corresponding author\\
email: julio@fisica.edu.uy

\vfill
\eject

\centerline{\large\bf Abstract}

\bigskip

We have evaluated the rate at which the asteroid belt is losing material, and how it splits between macroscopic bodies and meteoritic dust. The mass loss process is due to the injection of asteroid fragments into unstable dynamical regions, associated to mean-motion resonances with Jupiter, Saturn and Mars or secular resonances, from where they are scattered either to the region of the terrestrial planets or to the vicinity of Jupiter's orbit. Asteroid fragments that do not escape from the belt are ground down by mutual collisions to meteoritic dust. Under the assumption that 25\% of the zodiacal dust mass is of asteroidal origin, we find that the asteroid belt is currently losing a fraction of about $\mu_o \simeq 8.8 \times 10^{-5}$ Ma$^{-1}$ of its collisionally-active mass (without the primordial objects Ceres, Vesta and Pallas), about 20\% as macroscopic bodies, and 80\% as dust particles that feed the zodiacal dust cloud. Extrapolation of the current mass loss rate to the past suggests only a moderate increase of the asteroid belt mass and the mass loss rate around 3.0 - 3.5 Ga ago (by about 50\% and a factor of two respectively). Yet, should the computed $\mu_o$ be somewhat underestimated owing to the different uncertainties associated to its computation, the extrapolation to the past would lead to quite different results. For instance, a moderate increase in the computed $\mu_o$, say by a factor of three, would lead to an exponential increase of the asteroid mass and mass loss rate about 3.5 Ga ago. A greater asteroid mass loss rate in the past should be correlated with a more intense impact rate of the Earth, Moon and the other terrestrial planets, which is indeed what suggests the geologic record \citep{Hart07}.

\bigskip


\vfill
\eject

\section{Introduction}

Collisions of extraterrestrial bodies have played a key role in the origin and development of life on Earth as suppliers of volatile and organic material (e.g. Shoemaker 1983). Collisions may have also been responsible for mass extinctions that punctuated the development of life on Earth \citep{Alva80}. It is very likely that the asteroid belt has been by far the main source of projectiles once that the solar system settled in its current architecture and the population of leftover planetesimals was depleted. If this was the case, the impact rate of extraterrestrial bodies with the Earth and the other terrestrial planets should be correlated with the progressive loss of material from the asteroid belt. The current asteroid belt population is probably the relic of a much more substantial population of planetesimals and embryo planets that existed when the solar system formed. Its overall mass could have been as large as $\sim 1$ M$_{\oplus}$ \citep{Chap75,Bott05b}. Since the very beginnings the asteroid belt was crisscrossed by several two-body and three-body mean-motion resonances (MMR) with Mars, Jupiter and Saturn, and secular resonances that created chaotic zones \citep{Migl98,Nesv98}. Bodies and fragments created by mutual collisions that fell into these chaotic zones experienced a fast increase of their eccentricities to the point that they become planet-crossers and were rapidly removed from the asteroid belt.

The different size distribution of near-Earth asteroids (NEAs) compared to that of the main asteroid belt (e.g. Strom et al. 2005) indicates that the dynamical transport process must be size-dependent, favoring the injection of smaller asteroids into the inner solar system. The collisions among large asteroids produce asteroid families with myriad of fragments of different sizes that shared similar orbits, a feature that was noticed and studied by \citet{Hira18}. The orbits of the fragments will slowly depart from that of the parent asteroid, the smaller the fragment, the larger the shift rate. The main mechanism to shift the semimajor axis of small asteroids is the nongravitational force due to the anisotropic thermal radiation of bodies unevenly heated by the Sun's radiation, specifically the Yarkovsky and Yarkovsky-O’Keefe-Radzievskii-Paddack (YORP) effects \citep{Fari99}.

The injection of fragments into chaotic zones is not the only way by which matter is lost from the asteroid belt. Mutual collisions among asteroids produce meteoritic dust (sub-mm size or smaller) that evolves under the Poynting-Robertson drag and the solar radiation pressure. In this paper we will analyze the mass loss rate of the asteroid belt, both as macroscopic bodies and as meteoritic dust. Since our interest is also to analyze the mass depletion of the asteroid belt in parallel with the history of the impact rate of the Earth and the Moon, we will start addressing this latter issue.

\section{Geologic evidence of the past impact rate in the Earth and the Moon}

Craters on the surfaces of solid bodies are the records of past impacts. The crater coverage on the Earth's surface is quite inhomogeneous and concentrates on cratons (stable portions of the continental crust), so the studied impact structures only represent a small fraction of the overall Earth's surface (about 6\%). Unfortunately, erosion and plate tectonics have erased the oldest impact structures on Earth. This is not the case of the Moon where some basins are nearly 4 billion years (Ga) old and some lunar rocks are probably older than this. But let us first analyze what can we learn from the Earth and next we will pay some attention to the Moon and the terrestrial planets record. 

The best way to learn about the oldest megaimpacts on the Earth is through the spherule layers in the Earth's stratigraphy that such impacts produce (for the time being Earth is the only planet in which this geologic structures have been studied). A body hitting the Earth vaporizes a mass of target rock that expands and condenses into molten droplets called {\it spherules}. The diameter of the impactor can be related to the spherule layer deposited on the surface through the empirical equation \citep{John12,John16}.

\begin{equation}
  D (km) = 17(l_r/\xi)^{(1/3)},
\end{equation}  
where $\xi$ is the efficiency factor that varies between 0.5 and 2, and $l_r$ is the reduced layer thickness (in cm) defined as $l_r = 2 f_{sp}l$, where $l$ is the measured layer thickness and $f_{sp}$ the volume fraction of spherules in the layer. In the Earth's geological stratigraphy several layers of glass spherules have been discovered stretching between 2 and 3.5 Ga ago suggesting the occurrence of large impacts with a frequency from a few to some tens times greater than the current frequency (e.g. Johnson and Mellosh 2012).

The Moon, Mars and Mercury possess heavily cratered surfaces that provide us insights on the impact history of the terrestrial planets. In particular, from the size distribution of impact craters, the size distribution of the impactors can be inferred. \citet{Stro05,Stro15} found that the terrestrial planets and the Moon have been impacted by two different populations of objects: Population 1, that dominated at early times and was associated with a much higher impactor flux than Population 2 that dominated at later times after the formation of lunar maria up to the present. According to \citet{Stro05,Stro15}, the Population 1 impactor size-distribution matches very well that for the main-belt asteroids (MBAs), so these impactors were very likely ejected from their source region in a size-independent manner, possibly by mutual gavitational interactions, or resonance-sweeping in case that the Jovian planets were still in the migration process \citep{Mint09}. On the other hand, the good match of the size-distribution of Population 2 with that of the near-Earth object (NEO) population suggests that the latter is responsible of the projectiles that dominated for the last 3.8-3.7 Ga \citep{Stro05}.

An important aspect of the collisional history of the Earth, Moon and the other terrestrial planets is whether a ''terminal cataclysm'' or ``late heavy bombardment'' (LHB) occurred about 3.9 Ga ago. This episode is described as an spike in the impact cratering record (it lasted for $\sim 170$ million years (Ma)) that followed an early period of $\sim 0.5$ Ga characterized by a much less frequency of impacts. The evidence for such a cataclysm came from the observed Pb-U fractionation in lunar rocks, collected by the Apollo and Luna missions, whose ages were found to cluster at $\sim 3.9$ Ga pointing to a widespread shock metamorphism in the ancient lunar crust at that time \citep{Tera73,Tera74a,Tera74b}. This time corresponds to the formation of multi-ring impact basins like Serenitatis and Imbrium. As said above, the LHB theory argues that this episode was preceded by a smaller impact rate. In support of this theory \citet{Ryde90} argued that Apollo impact melt rocks dates clustered at $\sim 3.9$ Ga, while there is a lack of older impact melt rocks which should be taken as an evidence of a sudden increase of the impact cratering rate after a period of $\sim 0.5$ Ga of quiescence. 

However, the theory of a LHB or terminal cataclysm has recently been challenged by new evidence suggesting a much smoother distribution of impacts with a declining tendency covering the first Ga of the solar system ($\sim 4.5-3.5$ Ga ago) \citep{Zell17,Hart19}. For instance, impact melts and shock-altered ages among ordinary chondrites and Vesta-related Howardite-Eucrite-Diogenite (HED) achondrites show broad distributions of impact ages spanning between $\sim 4.5$ Ga to $\sim 3.5$ Ga rather than a sharp peak at 3.9 Ga. Impact melt clasts and other possibly impact-related materials in lunar samples were found to have radiometric ages (U-Pb, $^{40}$Ar-$^{39}$Ar) clustered around $\sim 4.21$ and $\sim 4.33$ Ga, i.e. much earlier than the putative LHB. These conflicting results have led to suggestions of basin-scale impacts at those earlier times, contrary to the terminal cataclysm/LHB paradigm invoking a relatively calm period during the first 0.5 Ga of the solar system lifetime.

We then assume here that the Earth-Moon system (and the other terrestrial planets) were subject to an intense and fast-declining bombardment as the reservoirs of projectiles (leftover planetesimals and the asteroid belt) spent most of their supply. \citet{Hart19} states that the cratering rate on the Moon 3.8 Ga ago was 150-200 times the present rate and could have reached 1000 times the present rate before $\sim 4.0$ Ga. It is very likely that in the first few hundreds Ma leftover planetesimals dominate the impact flux of the terrestrial planets \citep{Morb18,Nesv23}, but asteroids took over as the main source of projectiles once the population of leftover planetesimals declined. The exponential falloff in the bombardment flux was followed by a more quiescent stage of slow decline in the impact bombardment until present. The tail-end of the exponential falloff occurred around 2.5 Ga. Since the main source of projectiles after $\sim 3.5$ Ga is thought to be the asteroid belt, the variation in the impact rate of the terrestrial planets and the Moon should be correlated with its depletion. 

The goal of this paper is just to estimate how much mass is losing the asteroid belt at present, to extrapolate this result to the past, and to compare the curve of the temporal mass loss rate with that of the impact cratering rate derived from the geologic record. By the reasons given above, we will essentially stop our analysis at 3.5 Ga, since earlier leftover planetesimals may have played an increasing role as source of projectiles as we approach the birth of the solar system. The consideration of this early stage is out of the scope of this paper.       

\section{Mass loss from the inner and middle asteroid belt}

The main asteroid belt can be described as a collisional system that evolved to a power-law size distribution as the equilibrium configuration \citep{Dohn69}. Yet, this simple power-law is valid only for a self-similar system. Any discontinuity in the system will introduce departures from a linear relation (in a log-log diagram) that will show up  as a wavy pattern \citep{Camp94}. This is indeed what is observed, though we can still fit a power-law size distribution within a certain size range. According to \citet{Ivez01} the differential size distribution can be expressed as $(dN/dt) \propto D^{-(s+1)}$ where $s=3$ in the size range $5 < D(\mbox{km}) < 40$, and then flattens ($s=1.3$) down to 0.4 km. This size distribution reflects a relative high number of tens-km size asteroids as compared to km-size asteroids, in contrast with what is observed in NEAs (very few tens-km size asteroids versus a high number of km-size ones). The reason of this discrepancy will be explained below.

For the purpose of this study we can consider three zones within the asteroid belt: inner, middle, and outer belt whose two inner boundaries are the MMRs with Jupiter: 3:1 ($a=2.5$ au) and 5:2 ($a \simeq 2.83$ au). Asteroids that diffuse from the inner and middle belt will usually end up as NEAs before colliding with the Sun; asteroids that diffuse from the outer belt will be mostly ejected by Jupiter, with a large fraction of them having brief incursions in the NEA region before ejection. In this section we will focus on those asteroids diffusing from the inner and middle asteroid belt, that provide the great majority of the NEA population.  

\subsection{Large NEAs and Mars-crossers with $D > 10$ km}

The big fragments (sizes $D \gsim 10$ km) produced in catastrophic collisions in the asteroid belt are more difficult to move to unstable dynamical regions, so those that are finally scattered are much scarcer than their smaller counterparts. Their size or absolute magnitude ($H$) departs from the continuous $H$-distribution of NEAs of sizes smaller than $D \sim 10$ km (see, e.g. Harris and D'Abramo 2015; Granvik et al. 2018). Usually, the physical parameter that is directly determined from photometry is $H$ that can be related to the body's diameter $D$ if we know or guess its (visual) geometric albedo $p_v$ from, for instance, its taxonomic type. For conversion of absolute magnitude to diameter we use the formula (e.g. Harris and D'Abramo 2015)

\begin{equation}
  D = \frac{1329 \mbox{ km}}{\sqrt{p_v}}10^{-\frac{H}{5}}.
\end{equation}
For $D=10$ km we have $H=12.75$ for an average visual geometric albedo $p_v = 0.14$ suitable for NEAs \citep{Wrig16}.

At present we have three NEAs with $D>10$ km: (433) Eros with $D=16.8$ km, (1036) Ganymed with $D=37.7$ km and (4954) Eric with $D=10.8$ km. There is a fourth body: (3552) Don Quixote with $D=19.0$ km whose source region is controversial. It could come from the outer asteroid belt or Hildas, though it might be as well a Centaur coming from the transneptunian belt \citep{Momm14}. In this regard, the observation of some gaseous (CO$_2$) activity \citep{Momm14} might be a strong indication that Don Quixote come from the outer asteroid belt or even farther out. It is classified as of taxonomic type D whereas the three former bodies are of taxonomic type S. In order to analyze their dynamical evolution, we have performed numerical integrations for $\pm 30$ Ma (in the past and in the future) in a heliocentric frame of reference by means of the numerical integrator EVORB \citep{Fern02}, considering the perturbations of seven planets (from Venus to Neptune, while the mass of Mercury was added to that of the Sun). The accuracy parameter was $10^{-12}$. Nongravitational forces were not included in the computations. Our integrations show that Eros, Ganymed and Eric have dynamical lifetimes of at least a few $10^7$ yr. On the other hand, Don Quixote has a much shorter dynamical lifetime of about several $10^5$ yr and spends only brief periods in the NEA region. We will ignore Don Quixote for the moment and will come back to it when we consider the population of bodies removed from the outer asteroid belt.

In order to keep the population of large NEAs in steady-state we will require a rate of injection in the NEA region of the order $N_{large}/\tau_{large}$, where $N_{large}=3$ is the number of NEAs with $D>10$ km at present, and $\tau_{large}$ their typical dynamical lifetime, where $\tau_{large} \sim$ a few $10^7$ yr. Thus, we will require an injection rate of the order of one large NEA every $\sim 10$ Ma. Of course, this is a small-number statistics and thus very uncertain. In order to improve the statistics we have to resort to the sample of Mars-crossing asteroids (MCAs) that can be taken as the precursors of NEAs. In effect, they are evolving dynamically so sooner or later their perihelion distances will reach the NEA region before colliding with the Sun or being ejected hyperbolically \citep{Fern23}.

Table 1 shows the MCAs with $D > 10$ km whose dynamical lifetimes are found to be $\tau_{dyn} < 50$ Ma \citep{Fern23}. We also included (3178) Yoshitsune that is not currently a Mars-crosser but it will enter into this category very soon. The integration of its orbit with the code EVORB shows a fast evolution, becoming a NEA in less than one Ma. The first five objects of Table 1 very likely come from the inner or middle asteroid belt ($a \lsim 2.9$ au), whereas the last two are likely members of the outer belt or belong to the Centaur population. We will leave the last two bodies aside for the moment, and concentrate on those asteroids that are likely to come from the inner or middle asteroid belt.

\begin{table}[h]
\centerline{Table 1. Mars-crossers with $D>10$ km and $\tau_{dyn} < 50$ Ma}
\begin{center}
\begin{tabular}{|lllll|} \hline
object & Sp Type & $a$ (au) & $i$ (deg) & $\tau_{dyn}$ (Ma)$^{(*)}$ \\ \hline
(391) Ingeborg & S & 2.32 & 22.9 & 48.8 \\
(512) Taurinensis & S & 2.19 & 8.7 & 21.9 \\ 
(1204) Renzia & S & 2.26 & 2.45 & 43.1 \\
(2064) Thomsen & S & 2.18 & 6.1 & 24.8 \\
(3178) Yoshitsune & - & 2.72 & 6.8 & 0.93 \\
(102528) 1999US3  & - & 2.77 & 28.9 & 0.20\\
(301964) 2000 EJ37 & - & 4.62 & 10.1 & 0.15 \\ \hline
(*) From \citet{Fern23} & & & & \\
\end{tabular}
\end{center}
\end{table}

In order to improve the rather poor statistics of asteroids larger than 10 km, we combine the currently observed three largest NEAs and the five MCAs of Table 1 mentioned before. In doing so we come up with a transfer rate to the NEA region of about one of these objects every $\sim 10$ Ma, from which about 80\% will move on low-$i$ orbits ($<15^{\circ}$). The previous result is still rather uncertain (probably by a factor of two) but stll meaningful for our study. As we will see below, under reasonable assumptions, the mass loss involved in asteroids with sizes $>10$ km represents only a minor fraction of the total mass loss.

The computation of the total or cumulative mass of asteroids with $D > 10$ km removed from the asteroid belt is quite complex given their low number and uncertain upper size limit. The observations show that the largest asteroid that reached the NEA region has $D \sim 38$ km, while the largest MCA of Table 1 is (512) Taurinensis with $D=23.1$ km (cf. Fern\'andez and Helal 2023) so we can take $D \sim 40$ km as a suitable upper limit for asteroids removed from the belt. Furthermore, we will adopt an exponent $s=3$ for the differential size distribution $dN/dD$ of these bodies, i.e. the same as they show in the asteroid belt in the size range $5 \lsim D \mbox{( km)} \lsim 40$ \citep{Ivez01}. The total mass of large NEAs in the size range $D_1 < D < D_2$, assuming a population in steady-state, is for the particular case of $s=3$  (see eq.(A7b) of Appendix)

$$M(D_1<D<D_2) = C \left(\frac{\pi\rho}{6}\right)\ln{\left(\frac{D_2}{D_1}\right)}. \mbox{\hspace{1cm} (3a)}$$

In the general case that $s \neq 3$ we have instead of eq.(3a) the following equation

 $$M(D_1<D<D_2) = \frac{C}{3} \left(\frac{\pi\rho}{6}\right)^{s/3}\left[\frac{M_2^{1-s/3}-M_1^{1-s/3}}{(1-s/3)}\right], \mbox{\hspace{1cm} (3b)}$$
where $M_1 = (\pi/6)D_1^3\rho$ and $M_2 = (\pi/6)D_2^3\rho$, $\rho$ is the bulk density of the asteroids, and $D_1=10$ km, $D_2=40$ km are the lower and upper limit of the size range. $C$ is a constant related to the steady-state number of observed NEAs with $D > 10$ km: $N(D>10) \simeq 3$. It is given by

$$C = \frac{sN(D>10)}{\left(D_1^{-s}-D_2^{-s}\right)}.$$
We will adopt an average $\rho=0.25\rho_C + 0.75\rho_S = 0.25 \times 1.5 + 0.75 \times 2.5 = 2.25$ g cm$^{-3}$ which reflects an adequate mixture of taxonomic classes C and S asteroids, with bulk densities $\rho_C \sim 1.5$ g cm$^{-3}$, $\rho_S \sim 2.5$ g cm$^{-3}$, coming from the inner and middle asteroid belt \citep{Deme14}. By introducing the numerical values in eq.(3a) we obtain

$$M(10<D<40) = 1.49 \times 10^{16} \mbox{ kg}.$$

If $\tau_{dyn}$ is the typical dynamical lifetime of the large NEAs, that we estimate $\tau_{dyn} \sim 30 $ Ma for the three bodies of our sample, the mass loss rate under the form of large bodies ($D>10$ km) will be: $\dot{M}(10<D<40)=M(10<D<40)/\tau_{dyn}$. By introducing the numerical values we obtain

$$\dot{M}(10<D<40) \simeq  5 \times 10^{14} \mbox{ kg Ma$^{-1}$}. \mbox{\hspace{1cm} (4)}$$  
If we change the upper size limit for the large NEAs, for instance if we adopt $D_2=50$ km instead of $D_2=40$ km, we would get a mass loss rate only about 15\% greater. If the power-law size distribution of large NEAs were flatter with an exponent, for instance 2.3 (closer to that of smaller NEAs) instead of $s=3$, the mass loss rate would increase only by a factor of about two. We can also play with other values of $\tau_{dyn}$ leading to results that differ from the previous one by a factor of 2-3.

\subsection{NEA population with $D < 10$ km}

Smaller NEAs with $D<10$ km are more common and can be well represented by a cumulative luminosity function (CLF) of the type $\log{N_H(<H)} = a + \alpha H$ where $a$ and $\alpha$ are fitting parameters (e.g. Harris and D'Abramo 2015; Schunov\'a-Lilly et al. 2017, Granvik et al. 2018). From Granvik et al. (2018) (see their Fig. 20), we can split the CLF of NEAs in four branches within the $H$-ranges: 1) $12.75<H<16$ with a fitted slope $\alpha_1=0.465$ and a number of asteroids within this range estimated to be $N(12.75<H<16)=150$; 2) $16<H<19$ with $\alpha_2=0.42$ and $N(16<H<19)=2695$; 3) $19<H<23$ with $\alpha_3=0.322$ and $N(19<H<23)=52739$; 4) $23<H<30$ with $\alpha_4=0.558$ and $N(23<H<30)=4.4463 \times 10^8$. We can easily convert a CLF into a CSD of the form $f(>D) \sim D^{-s}$ bearing in mind that $s=5\alpha$ (see Appendix). We get the corresponding exponents $s_i$ ($i$=1,4), which are valid within the size ranges: 1) $2.24 < D(\mbox{km}) < 10$, $s_1=2.325$; 2) $0.563 < D(\mbox{km}) < 2.24$, $s_2=2.1$; 3) $0.0892 < D(\mbox{km}) < 0.563$, $s_3=1.61$; 4) $3.55\times 10^{-3} < D(\mbox{km}) < 0.0892$, $s_4=2.79$.

For all the values $s \neq 3$ we can apply eq.(3b) to compute the integrated mass $M(\Delta D i)$ of NEAs within the four $H$ (or $D$) ranges given above, which leads to the results shown in Table 2.

\begin{table}[h]
\centerline{Table 2. Masses of NEAs within given size ranges}
\begin{center}
\begin{tabular}{|ll|} \hline
Size range (km) & Mass (kg) \\ \hline
$2.24<D<10$ & $1.23 \times 10^{16}$ \\
$0.563<D<2.24$ & $3.45 \times 10^{15}$ \\
$0.0892<D<0.563$ & $6.44 \times 10^{14}$ \\
$0.00355<D<0.0892$ & $3.02 \times 10^{14}$ \\ \hline
\end{tabular}
\end{center}
\end{table}

By summing the masses of the four size ranges of Table 2, we get the total mass of NEAs:

$$M(0.00355<D<10) =  1.67 \times 10^{16} \mbox{ kg}.$$

The NEA population will be lost in a time scale that depends on its source region. There are four important source regions in the inner and middle asteroid belt: 1) Hungaria, 2) $\nu_6$ secular resonance, 3) Phocaea, and 4) 3:1 mean motion resonance. According to \citet{Gran18}, the fraction $\gamma$ of asteroids supplied by each one of these regions and their dynamical lifetimes $\tau$ are respectively: 1) $\gamma_{Hun}=0.02$ and $\tau_{Hun}=73.2$ Ma; 2) $\gamma_{\nu 6}=0.07$ and $\tau_{\nu 6}=7.5$ Ma; 3) $\gamma_{Pho}=0.03$ and $\tau_{Pho}=11.2$ Ma; 4) $\gamma_{3:1}=0.2$ and $\tau_{3:1}=2.0$ Ma. The weighted average dynamical lifetime of NEAs will be $\bar{\tau}_{dyn} = 6.3$ Ma (remember that this estimate does not take into account asteroids coming from the outer asteroid belt). Therefore, the mass loss rate of NEAs smaller than $D=10$ km will be given by 

$$\dot{M}(0.00355<D<10) = \frac{M(0.00355<D<10)}{\bar{\tau}_{dyn}} = 2.65 \times 10^{15} \mbox{ kg Ma$^{-1}$}. \mbox{\hspace{1cm} (5)}$$

By comparing eqs.(4) and (5) we see that the bulk of the mass loss is due to bodies with $D < 10\mbox{ km}$. Despite the danger posed by the big asteroids with $D>10$ km of a catastrophic collision with the Earth, their contribution is a minor fraction (about 16\%) of the whole mass loss rate of all NEAs. As we have seen before, this is in contrast with what is observed in the asteroid belt where asteroids with sizes above 10 km and up to several tens km contain more mass than their smaller counterparts. The explanation is that given their large sizes is more difficult to move them to dynamically unstable zones of the asteroid belt by forces associated to the thermal radiation (Yarkovsky and YORP effects), so they will tend to stay longer in the asteroid belt.

\subsection{Total mass loss from the inner and middle asteroid belt}
 
Summing the contributions in all the size ranges we obtain for the mass loss rate associated with bodies coming from the inner and middle asteroid belt:

$$\dot{M}_{inner+middle} \simeq \dot{M}(10<D<40) + \dot{M}(0.00355<D<10) \simeq 3.15 \times 10^{15} \mbox{ kg Ma$^{-1}$}. \mbox{\hspace{1cm} (6)}$$

We notice that the size distribution is truncated at $D=3.55$ m. This limit has little relevance as smaller particles will have a negligible mass contribution to the NEA population. At such small sizes, the bodies will be subject to frequent catastrophic collisions that will finally grind down the boulders to meteroritic dust that will be swept by the solar radiation forces (Poynting-Robertson drag and radiation pressure). \citet{Fari98} estimate the collisional lifetime of a silicate asteroid as

$$\tau_{coll} = 2.0 \times 10^7 \mbox{ yr}\left(\frac{R}{1\mbox{ m}}\right)^{1/2}. \mbox{\hspace{1cm} (7)}$$

The results obtained from eq.(7) within the size range $10^{-4} \lsim D(\mbox{km}) \lsim 1$ are in fairly good agreement with those obtained by \citet{Bott05b} from a collisional and dynamical depletion model. Yet for sizes greater than $\sim 10$ km, the collisional lifetimes found by \citet{Bott05b} exceed the age of the solar system. At any rate, asteroids with diameters $D \lsim 10$ km will be subject to a continuous collision process leading to a progressive large number of smaller fragments on a time scale shorter than the solar system age. The collisional lifetime of submeter boulders will be of a few Ma. The final product of this process will be micron-size meteoritic dust that will be incorporated into the zodiacal cloud. We will discuss below how much mass is lost from the asteroid belt as dust.

\section{Mass loss from the outer asteroid belt}

Most of the asteroids in the outer belt are concentrated in the range of semimajor axes $2.8 \lsim a(\mbox{au}) \lsim 3.28$, namely between the 5:2 and 2:1 MMRs with Jupiter. It is striking the falloff of the asteroid population beyond the 2:1 MMR that may be an imprint of the early planet migration, in particular Jupiter's. Numerial simulations carried out by \citet{Liou97} showed that the inward migration of Jupiter in the early solar system was responsible fot the dynamical destabilization and removal of asteroids in the range $3.5 \lsim a(\mbox{au}) \lsim 3.9$. The outer asteroid belt is crisscrossed by several MMRs with Jupiter, such as the 7:3, 9:4, 2:1, 13:7, 9:5 and 7:4 (e.g. Bottke et al. 2002; Gallardo 2006) and also several three-body mean-motion resonances (Jupiter-Saturn-asteroid) (Nesvorn\'y and Morbidelli 1998) which turns the region as highly chaotic. Chaotic diffusion in eccentricity will place outer-belt asteroids in orbits close to Jupiter leading to their fast removal from the planetary region. Furthermore, the outer asteroid belt is the place of several large families such as Eos, Hygiea and Themis. In particular the Veritas family, close to the inner side of the 2:1 MMR with Jupiter, may have formed in the past 50 Ma \citep{Mila94}.

According to the model results by \citet{Bott02}, $\sim 85\%$ of the NEOs come from the inner and middle asteroid belt, and $\sim 8\%$ from the outer asteroid belt ($a > 2.83$ au) (the rest comes from the Jupiter-family comet region). \citet{Bott02} also estimate that bodies coming from the outer asteroid belt spend in NEO orbits on average $\sim 0.14$ Ma. If we bear in mind that NEAs coming from the inner and middle asteroid belt have an average dynamical lifetime of 6.3 Ma (cf. Section 3.2), we find a ratio of asteroids removed from the outer belt to the inner + middle belt of: $N(\mbox{outer})/N(\mbox{inner$+$middle})=(0.08/0.85) \times (6.3/0.14) \simeq 4.2$. \citet{Nesv18} found that about 5 times more large asteroids ($D>10$ km) leak into NEA orbits from $a > 2.9$ au than from $a < 2.9$ au, result that is consistent with the previous one. On the other hand, they also found that the number of large main-belt asteroids with $a>2.9$ au is only about a factor 2.3 greater than those with $a < 2.9$ au. These results suggest that the loss of material from the outer belt is about twice that of the inner and middle belt, at least in what concerns asteroids with $D>10$ km. Is this due to a circumstantial event (for instante a recent catastrophic collision that generated a myriad of fragments)?, or is it a permanent feature related to the dynamics in the outer belt that leads to a greater fraction of mass loss in comparison with the inner and middle belt? This is an interesting question that requires further study.

\begin{figure}
\includegraphics[width=0.6\textwidth]{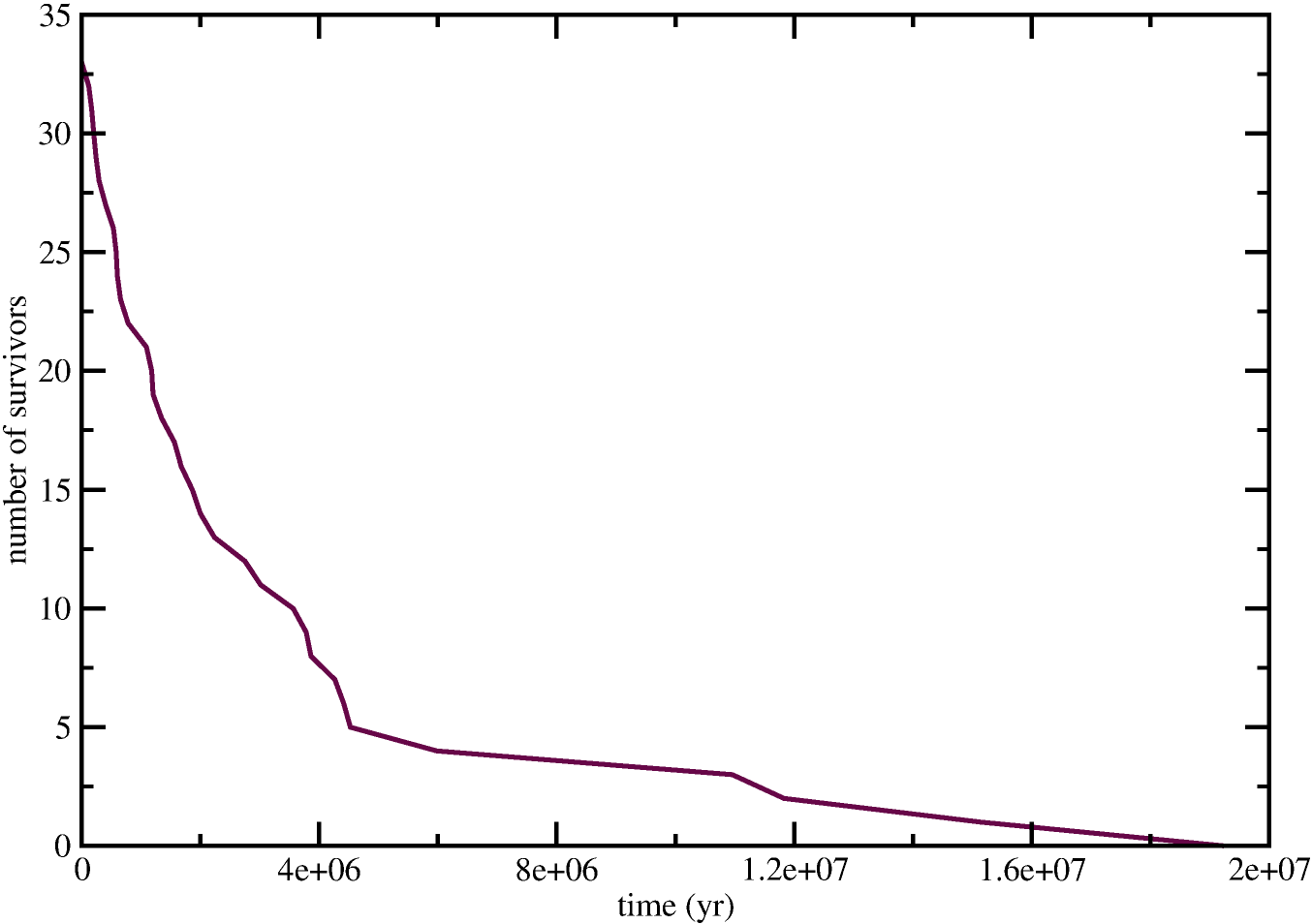}
\caption{The decrease of the Mars-crossing population with semimajor axes $2.9 < a < 5.2$ au as a function of time. Most of the objects end up ejected in hyperbolic orbits by the strong gravitational perturbations of Jupiter.}
\label{mcas_a_gt_29}
\end{figure}

In order to check the previous results, we carried out our own estimate of the mass loss from the outer asteroid belt. To this end, we considered the population of MCAs with absolute magnitudes $H < 16$ and aphelion distances $Q < 4.5$ au as adopted by \citet{Fern23} for their simulations. These authors showed that the great majority of MCAs come from the Hungarias, the inner and middle main belt. They found that $\sim50\%$ of the high-inclination MCAs ($i \gsim 15^{\circ}$) remain in stable orbits after one Ga, while the percentage decreases to $\sim 8\%$ for the low-$i$ MCAs ($i \lsim 15^{\circ}$). The dynamical lifetime of the 652 MCAs that are effectively diffusing (namely, those that have dynamical lifetimes $<$ one Ga) is found to be $\tau_{dyn}(\mbox{inner+middle}) \sim 120$ Ma. As said before, the great majority come from the inner and middle asteroid belt with very little contamination from the outer belt, so we do not think that this contamination will affect the previous result. We compare the previous result with that derived for the more distant MCAs with $a > 2.9$ au which presumably are coming from the outer belt. In order to avoid the inclusion of objects in highly eccentric orbits that might have a comet origin in the transneptunian region, we set an upper limit for the semimajor axis $a < 5.2$ au. By considering the same magnitude limit $H=16$, we were left with a sample of 34 distant MCAs. In order to study their dynamical evolution, we integrated their orbits for 30 Ma with the EVORB code with similar specifications to those described in Section 3.1. At the end of the considered period, all the objects were removed from the asteroid belt with a dynamical half-life of $\tau_{dyn}(\mbox{outer}) \sim 1.7$ Ma. The results are shown in Fig. \ref{mcas_a_gt_29}. If we make allowance for the different dynamical half-lives, the ratio between the MCA populations removed from the outer asteroid belt and the inner+middle asteroid belt will be $N(\mbox{outer})/N(\mbox{inner+middle}) = 34/652 \times (120/1.7) \sim 3.7$. Very likely this ratio is a lower limit bearing in mind that the sample of MCAs with $a > 2.9$ au has a greater degree of incompleteness given their greater distances to the Sun (and the Earth). To try to gauge the incompleteness effect, we consider the restricted samples of MCAs brighter than $H=15$ (237 with $a<2.9$ au and 14 with $a > 2.9$ au), from which we obtain $N(\mbox{outer})/N(\mbox{inner+middle}) = 14/237 \times (120/1.7) \sim 4.2$. The previous results show consistently that the mass removed from the outer belt is about $\sim 4-5$ times greater than that coming from the inner+middle belt.

\section{Total mass loss from the asteroid belt as macroscopic bodies}

We come to the point of estimating all the mass that is being removed from the asteroid belt as macroscopic bodies, stretching through a wide range of sizes, from several tens km down to meter-sized meteoroids. 

$$\dot{M}_{macro} \simeq \dot{M}_{inner+middle} + \dot{M}_{outer} = \dot{M}_{inner+middle}\left(1 + \frac{\dot{M}_{outer}}{\dot{M}_{inner+middle}}\right). \mbox{\hspace{1cm} (8)}$$  

By substituting by the numerical results previously obtained and adopting the ratio $\dot{M}_{outer}/\dot{M}_{inner+middle} \simeq 4.5 \pm 0.5$, according to what was discussed in the previous section, we find

$$\dot{M}_{macro} \simeq (1.73 \pm 0.16) \times 10^{16} \mbox{ kg Ma$^{-1}$}. \mbox{\hspace{1cm} (9)}$$  

\section{Total mass loss from the asteroid belt as meteoritic dust}

Through mutual collisions asteroids with sizes $\lsim 30$ km are ground down to dust over time scales shorter than the solar system age. A collision cascade will set up leading to meteoroids and micron-size dust particles that will be under the influence of the Poynting-Robertson (P-R) drag and the Sun's radiation pressure. This is another way in which mass can be lost from the asteroid belt. This dust material will be removed from the asteroid belt and incorporated into the zodiacal cloud. \citet{Nesv10} estimate the mass of the zodiacal cloud within 5 au from the Sun at $1-2 \times 10^{16}$ kg, mainly in $D=100-200$ $\mu$m particles. The time of spiralling into the Sun by the P-R drag for a black, perfectly absorbing spherical particle of radius $s$ and bulk density $\varrho$ in an orbit of perihelion distance $q$ is \citep{Whip67}

$$\tau_{PR} = C(e)\varrho s q^2 \times 10^7 \mbox{ yr}, \mbox{\hspace{1cm} (10)}$$
where $s$ and $\varrho$ are in cgs units, $q$ in au, and $C(e)$ is a coefficient that depends on the orbit's eccentricity $e$. $C(e) \sim 1$ for low eccentricities as is the case of particles in the asteroid belt. As an example: a dust particle of $s=100$ $\mu$m, $\varrho=1$ g cm$^{-3}$, that starts at $q \sim 2.5$ au will require a time $\tau_{PR} = 6.25 \times 10^5$ yr to spiral into the Sun. In the meantime, the particle will suffer catastrophic collisions with a time scale (cf eq.(7)): $\tau_{coll} = 2 \times 10^4$ yr that will accelerate its final demise. \citet{Whip67} estimated a lifetime for zodiacal dust particles of $\sim 10^5$ yr, so that to keep the zodiacal cloud in equilibrium a mass input of $1-2 \times 10^{16}/10^5 \simeq 1-2 \times 10^{11}$ kg yr$^{-1}$, or 3.2-6.4 tons s$^{-1}$ was found to be required. 

\begin{figure}
\includegraphics[width=0.6\textwidth]{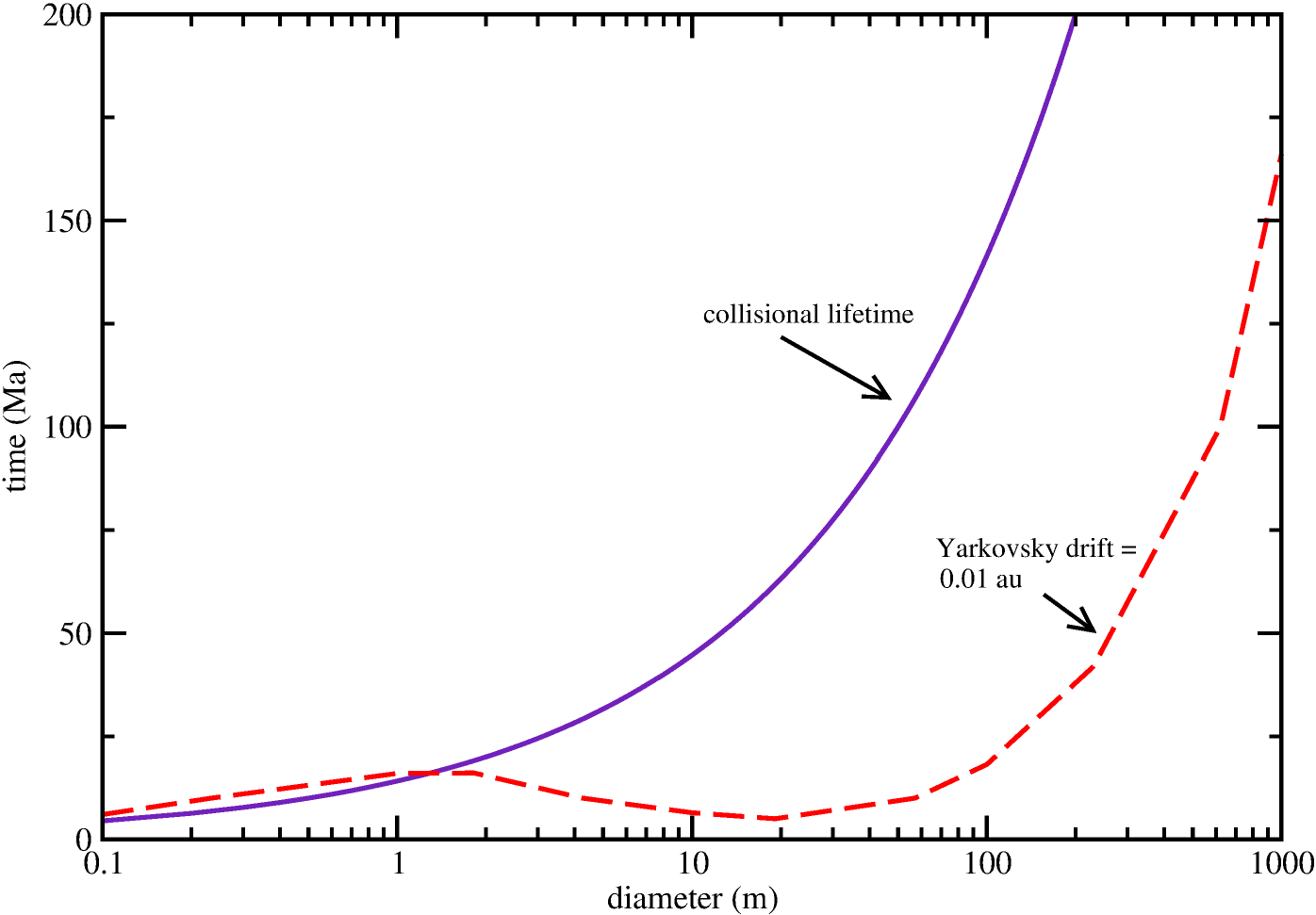}
\caption{Collisional lifetimes of bodies in the asteroid belt as a function of their diameters $D$ as derived from eq.(7) (solid curve). Time rquired for a body of diameter $D$ to experience a shift of its semimajor axis of $\Delta a = 0.01$ au by the Yarkovsky mechanism based on data from Fig. 3 of \citet{Broz06} (dashed curve).}
\label{yarkovsky_drift}
\end{figure}

At this point we can try to be more specific about what we understand as "macroscopic bodies" and "meteoritic dust". Let us assume that a body that experiences a radial shift in semimajor axis of $\sim 10^{-2}$ au$^{-1}$ will fall in a chaotic zone and be removed from the asteroid belt. If the time scale for this radial shift is shorter than the collisional lifetime we can say that the body will be removed intact as a macroscopic body. As shown in Fig. \ref{yarkovsky_drift} bodies larger than $D \gsim 1$ m will likely be removed from the asteroid belt before suffering a catastrophic collision. On the other hand, bodies smaller than $\sim 1$ m will continue their fragmentation cascade in the asteroid belt down to dust that will be finally removed by P-R drag. As bodies are ground down to tiny dust particles the Yarkovsky/YORP dependence with $D^{-1}$ will no longer hold as heat conduction in the interior of the particles will lead to a nearly isothermal surface and a nearly isotropic thermal emission \citep{Broz06}. When this point is reached the solar radiation pressure and the P-R drag will take over as the main responsible for moving particles, 

There are two main sources believed to supply most of the dust of the zodiacal dust complex: Jupiter-family comets and asteroids (e.g. Nesvorn\'y et al. 2010). There has been a lot of discussion about what is the contribution from asteroids in quantitative terms. Several studies on the collisional dust production of the asteroid belt as well as the nature of the interplanetary dust particles and meteoritic material that reach the Earth suggest that the contribution of asteroids to the zodiacal dust complex is significant. For instance from fireball observations, \citet{Shob21} estimate that as much as $94.2\% \pm 3.2\%$ of Earth-crossing centimeter-size meteoroids in cometary orbits come from the asteroid belt. Along the same lines, \citet{Zook86} also argue that asteroids contribute a major, possibly dominant, fraction of the interplanetary dust particles with radii less than 50 $\mu$m. The authors based their conclusion on the warm dust bands detected by IRAS in the asteroid belt, presumably the collision products of asteroids, and the nature of particles that impacted the thermal blanket of the Solar Max satellite which were found to be rich in Fe-Ni sulfides common in meteorites of asteroidal origin. From the analysis of the peak temperatures reached by dust material collected in the stratosphere, \citet{Flyn89} also concluded that most of this material comes from parent bodies in the main asteroid belt.

A sophisticated model developed by \citet{Dohn69}, based on the dynamico-collisional evolution of main belt asteroids led to an estimate of the amount of mass lost yearly from the asteroid belt of about $6 \times 10^{10}$ kg or about 2 tons s$^{-1}$. It is interesting to stress that this result corresponds to pure asteroid losses without taking into account the contribution from comets. \citet{Grun85} estimate that about 9 tons s$^{-1}$ of meteor-size particles ($m \gsim 10^{-5}$ g) are lost inside 1 au due to collisions, while the smallest fragments generated in the collisions  ($m \lsim 10^{-10}$ g) are blown away by the solar radiation pressure as $\beta$ meteoroids. \citet{Rowa13} derive an asteroidal contribution of 22\% from the modeling of the infrared emission from zodiacal dust detected by the IRAS and COBE missions. \citet{Rigl22} developed a model for the distribution of interplanetary dust by considering comet fragmentation as the main source. For their simulations they follow the dynamical evolution of fictitious Jupiter-family comets that are subject to random fragmentation that produce dust. They fit their computed dust distribution with some observed parameters of the zodiacal cloud such as the slope of the radial distribution, and the grain size which dominates the cross-sectional area at 1 au. The best fit is obtained with a mass input to the zodiacal cloud between 6.2 and 11.1 tons s$^{-1}$ in good agreement with that found by \citet{Grun85}.  

The previous studies tend to agree that the zodiacal cloud requires the input of several tons s$^{-1}$ (say 5-10) of dust material in order to keep it in steady state. The dust can be supplied either by comets or asteroids. There is a general agreement that comets are the main contributor, though the contribution from asteroids can be significant, from as low as about 10\% \citep{Nesv10} to more than one third \citep{Liou95,Durd97}.

We will assume in the following that a fraction $f$ of the dust mass deposited in the zodiacal cloud comes from asteroids. For instance, if a mass input rate of 9 tons s$^{-1}$ \citep{Grun85} is required to keep the zodiacal cloud in steady state, the asteroid belt will supply     

$$\dot{M}_{dust} \simeq f \times 9 \mbox{ tons s$^{-1}$} \simeq f \times 28.4 \times 10^{16} \mbox{ kg Ma$^{-1}$}. \mbox{\hspace{1cm} (11)}$$

If we adopt $f=0.25 \pm 0.1$ for the fraction of dust coming from the asteroid belt, we obtain

$$\dot{M}_{dust} \simeq (7.1 \pm 2,8) \times 10^{16} \mbox{ kg Ma$^{-1}$},$$
that corresponds to the mass loss rate from the asteroid belt as dust particles.

We note that our estimate closely agrees with \citet{Dohn69}'s theoretical computation of the rate of mass loss from the asteroid belt as dust particles generated through mutual collisions.

\section{The current rate of mass loss from the asteroid belt}

By summing the contributions from macroscopic bodies and meteoritic dust, we obtain the current mass loss rate from the main asteroid belt

$$\left(\frac{\Delta M}{\Delta t}\right)_o = \dot{M}_{macro} + \dot{M}_{dust} = (1.73 + 28.4f) \times 10^{16} \mbox{ kg Ma$^{-1}$}. \mbox{\hspace{1cm} (12)}$$

\begin{figure}
\includegraphics[width=0.6\textwidth]{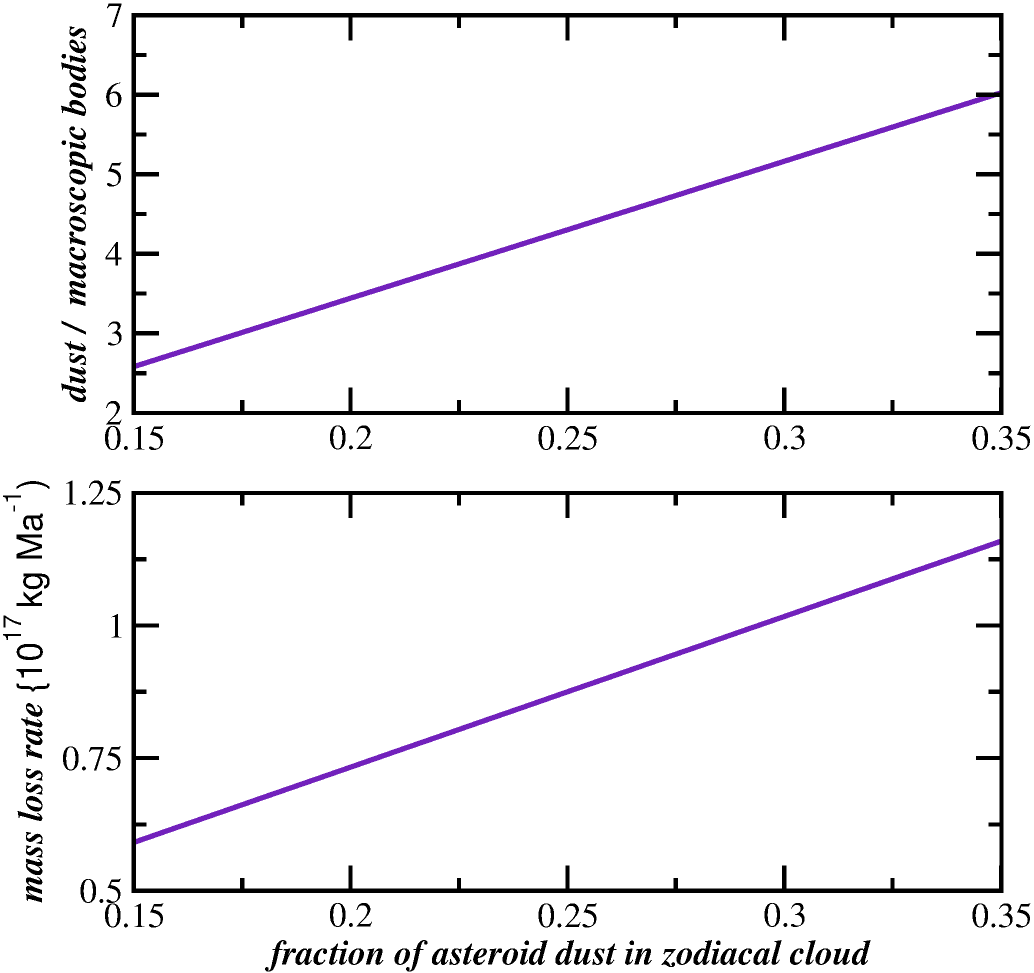}
\caption{Dust-to-macroscopic bodies mass loss rate ratio (upper panel) and total mass loss rate from the asteroid belt (lower panel) as a function of the fraction of dust in the zodiacal cloud assumed to come from asteroids.}
\label{massloss_f}
\end{figure}

By adopting $f=0.25$, we get

$$\left(\frac{\Delta M}{\Delta t}\right)_o \simeq 8.83 \times 10^{16} \mbox{ kg Ma$^{-1}$},$$
from which about 20\% is lost as macroscopic bodies and 80\% as dust particles. If $f$ takes other values, $\left(\frac{\Delta M}{\Delta t}\right)_o$ as well as the relative dust/macroscopic bodies contribution will change accordingly. In the upper panel of Fig. \ref{massloss_f} we plot the mass loss rate ratio of dust to macrocopic bodies as a function of $f$. We consider for $f$ a range of values between 0.15 and 0.35 that cover most of the estimates from several authors as discussed above. For the whole range of $f$, the dust component predominates, reaching more than 85\% of the total mass for $f=0.35$. The range of reasonable values of $f$ only affects the total mass loss rate by a factor of about 3.5 (see lower panel of Fig. \ref{massloss_f}).

Let $\mu_o$ be the current rate of relative mass loss from the main asteroid belt that is given by

$$\mu_o = -\frac{(\Delta M /\Delta t)_o}{M_o}, \mbox{\hspace{1cm} (13)}$$ 
where $M_o$ is the current mass of the collisionally evolved asteroid belt in steady-state that can be described by a power-law mass distribution \citep{Dohn69}. This criterion leaves aside the three largest asteroids (Ceres, Vesta, Pallas) that have remained more or less intact from the early solar system, and have masses detached from the rest of the asteroids. From dynamical models of the solar system and spacecrafts, \citet{Pitj18} derived a mass for the asteroid belt of $2.4 \times 10^{21}$ kg from which $1.43 \times 10^{21}$ kg corresponds to the three largest asteroids. We are then left with a mass under collisional evolution of $M_o \simeq 10^{21}$ kg. By introducing in eq.(13) the numerical values found before, we get:

$$\mu_o \simeq -8.83 \times 10^{-5} \mbox{ Ma$^{-1}$}, \mbox{\hspace{1cm} (14)}$$
where we assume that $f=0.25$.

\section{The mass loss from the asteroid belt through time: Theoretical estimate}

The idea that the current asteroid belt is the remnant of a much larger population is not new \citep{Weid75,Chap75,Weth92}. This is shown by the power-law size distribution of asteroids as the result of a collisionally evolved population, being the observed asteroid families the footprints of megacollisions between massive asteroids that occurred in the past \citep{Nesv15}. In order to extrapolate the current mass loss from the asteroid belt back into the past, we will follow a procedure similar to that developed by \citet{Chap75}. If we assume that the depletion of the $N(M)$ asteroids of masses within ($M,M+dM$) is caused by mutual collisions, then the depletion rate will be proportional to $N^2$, so it can be expressed as

$$\frac{dN}{dt} = -kN^2, \mbox{\hspace{1cm} (15)}$$
where $k$ is a constant. The effect of mutual collisions is twofold: 1) to produce fragments that reach dynamically unstable regions of the belt from where they are removed (loss as macroscopic bodies), and 2) to produce dust that is removed from the belt by P-R drag (loss as dust).

If we assume that the objects in the population have an average mass $\bar{m}$, so the mass of the $N$ asteroids is $M=N\bar{m}$, the previous equation can be converted into an equation for the mass loss from the asteroid belt of the form

$$\frac{dM}{dt} = -\frac{k}{\bar{m}} M^2, \mbox{\hspace{1cm} (16)}$$  
where the asteroid mass distribution and $\bar{m}$ change very little with time, so we can substitute $M/\bar{m}=N$ and $[(dM/dt)/M]_o =\mu_o$ in eq.(16) assuming $\bar{m}$ to be constant in time. The assumption of a nearly constant $\bar{m}$ gets support from numerical simulations of the collisional evolution of the main asteroid belt and the impact cratering record on the Moon, showing that the size distribution of asteroids have changed very little since the early solar system \citep{Bott05a,Stro05,Obri11}.

At present $M=M_o$ so we obtain

$$\mu_o = -\frac{k}{\bar{m}}M_o. \mbox{\hspace{1cm} (17)}$$

We can integrate eq.(16) between a time $t$ (in the past) and the present time $t_o$, and introduce eq.(17), leading to

$$M(t) = \frac{M_o}{1+{\mu}_o(t_o-t)}, \mbox{\hspace{1cm} (18)}$$
while the mass loss rate is

$$\dot{M}(t) = - \frac{M_o{\mu}_o}{[1+{\mu}_o(t_o-t)]^2}. \mbox{\hspace{1cm} (19)}$$
  
\begin{figure}
\includegraphics[width=0.6\textwidth]{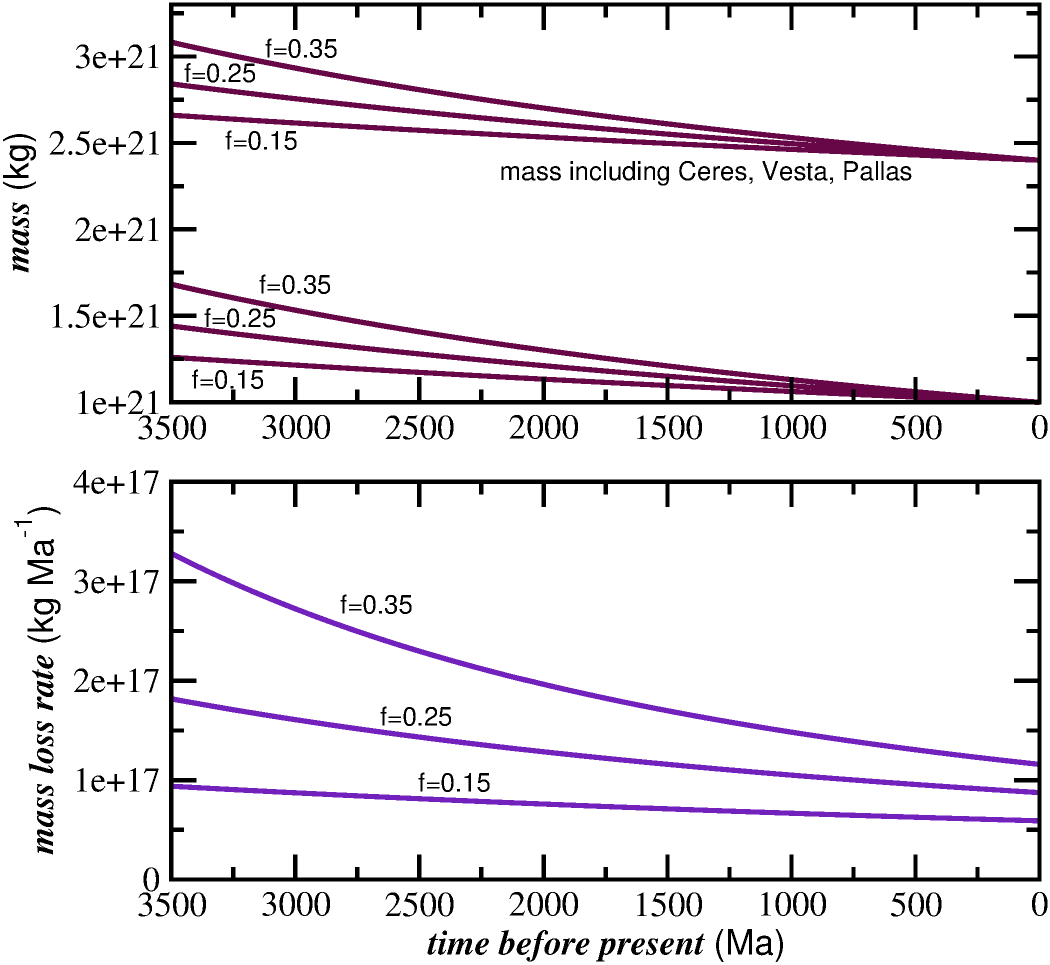}
\caption{Extrapolated mass to the past 3.5 Ga for three values of the fraction of asteroid mass contribution to the zodiacal cloud: $f=0.15$, $f=0.25$ and $f=0.35$. We consider the asteroid belt mass with and without the three largest asteroids Ceres, Vesta and Pallas (upper panel). Mass loss rate from the asteroid belt through time for the three values of $f$ considered above (bottom panel).}
\label{mass_loss}
\end{figure}

The computed mass of the asteroid belt and the mass loss rate are shown in Fig. \ref{mass_loss} as a function of time. The main asteroid belt could have been only about 50\% more massive around 3.5 Ga ago, and the mass loss rate about twice higher, if we adopt our canonical value of $\mu_o$ of eq.(14). These computed values seem to be rather low in comparison with some geologic evidence analyzed before suggesting a more massive asteroid belt and a concomitant more intense impact bombardment.

There is the possibility that the computed $\mu_o$ of eq.(14) has fluctuations over time scales of tens to a few hundreds Ma. It might happen that we are now in a low in the mass loss rate from the asteroid belt due to the lack of megacollisions in the recent past but the situation may be quite different over time scales of several hundreds Ma. For instance, \citet{Bott07} argued about an increase in the impact flux by a factor of two in the past $\sim 100$ Myr which was due probably to the catastrophic disruption of (298) Baptistina. \citet{Schm03} also found an enhancement of two orders of magnitude in the influx of meteorites on Earth in the mid-Ordovician ($\sim 460$ Ma ago) from the abundance analysis of chromite grains from decomposed chondritic meteorites. The source of this meteorite enrichment might have been the catastrophic disruption of the parent body that gave rise to the large Flora asteroid family about $\sim 480$ Ma ago. From the analysis of the thermophysical characteristics of the lunar impact ejecta, measured with the thermal radiometer of the {\it Lunar Reconaissance Orbiter} (LRO), \citet{Mazr19} were able to estimate ages of craters with diameters $>10$ km younger than one Ga. They found that the impact rate increased by a factor of 2.6 about 290 Ma ago. All these pieces of evidence coming from the geologic record of the Earth and the Moon suggest that fluctuations by factors of two-three in the impact cratering rate of the Earth and the Moon may be quite possible over time scales of a few hundreds Ma. And these fluctuations in the impact cratering rate should be correlated with the mass loss rate $\mu_o$ from the asteroid belt.

We have to point out that the computed value of $\mu_o$ given by eq.(14) has intrinsic error bars inherent to the many assumptions we made for its derivation. The main source of uncertainty is in our opinion the asteroidal contribution to the zodiacal cloud. As discussed before, our assumed fraction of $f=0.25$ may be greater or smaller according to different estimates, that usually fall within, say $f=0.15-0.35$. Therefore the value of $\mu_o$ may have an uncertainty by a factor of a few units, that may lead to different possible scenarios for the mass evolution and mass loss rate of the asteroid belt. Fig. \ref{mass_loss} shows how different fractions of the dust contribution from the asteroid belt to the zodiacal cloud ($f$ from 0,15 to 0,35) lead to important changes in the mass and mass loss rate of the asteroid belt in the past. The computed mass of the asteroid belt 3.5 Ga ago is found to vary by factors between 1.3 and 1.7 with respect to the current mass for $f=0.15$ and $f=0.35$ respectively, while the mass loss rate varies by factors between 1.6 and 3 with respect to the current one.

\begin{figure}
\includegraphics[width=0.6\textwidth]{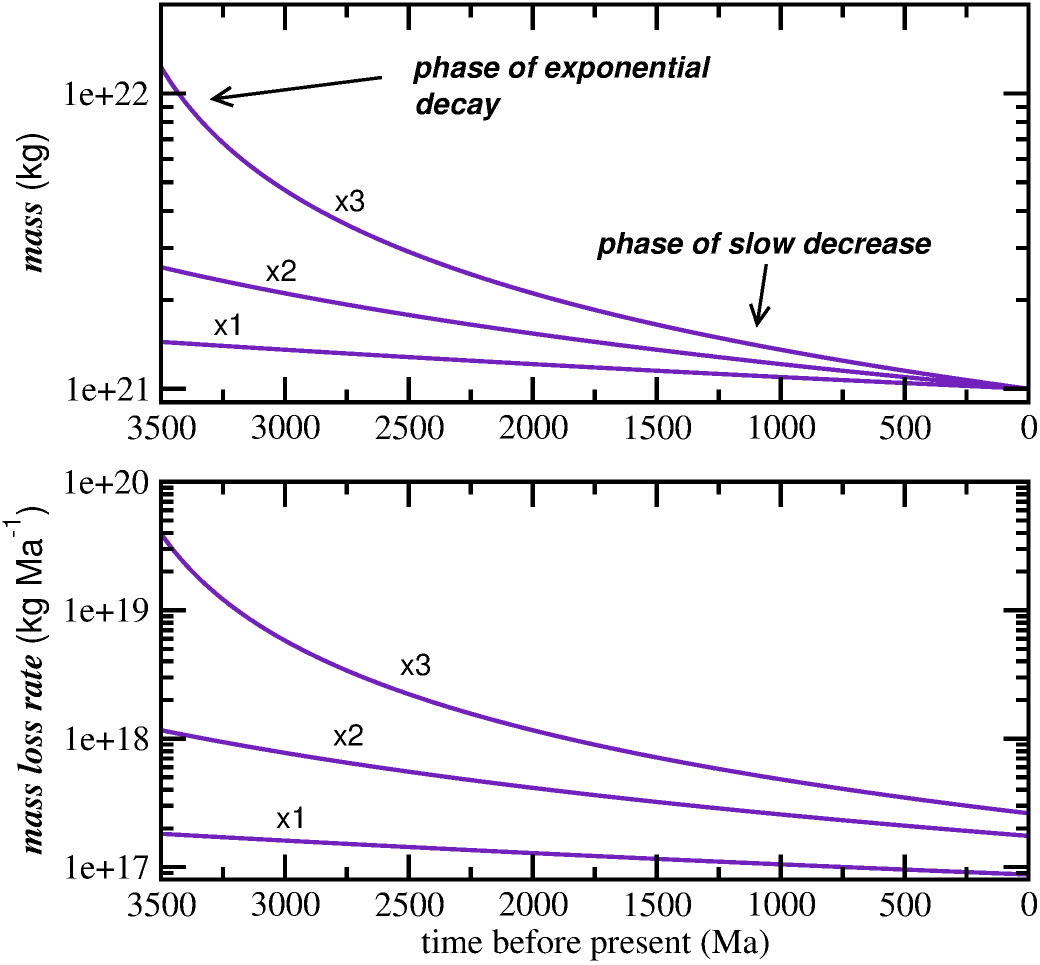}
\caption{Extrapolated mass to the past of the collisonally active asteroid belt (without Ceres, Vesta, Pallas) in the last 3.5 Ga assuming a relative mass loss rate $\mu_o$ at present equal to one, two, three times the value derived in eq,(14) (upper panel). The corresponding mass loss rate for the same conditions as before (bottom panel).}
\label{belt_evol_prim}
\end{figure}

\section{The primordial asteroid belt}

Fig. \ref{belt_evol_prim} shows different scenarios obtained by scaling the computed value of $\mu_o$ of eq.(14) by factors between 1-3. A trend to an ever increasing asteroid belt mass and mass loss rate in the distant past is observed as we apply larger factors to $\mu_o$. For $3 \times \mu_o$ there is an exponential increase of both asteroid mass and mass loss rate (and the correlated impact rate of the terrestrial planets) at about 3-3.5 Ga. If we increase our computed $\mu_o$ (or a time-average $\mu_o$ over a period of a few hundreds of Ma) by only a factor of two, it would lead to an increase in the mass loss rate by a factor of about 7.5 about 3.5 Ga ago.

\begin{figure}
\includegraphics[width=0.6\textwidth]{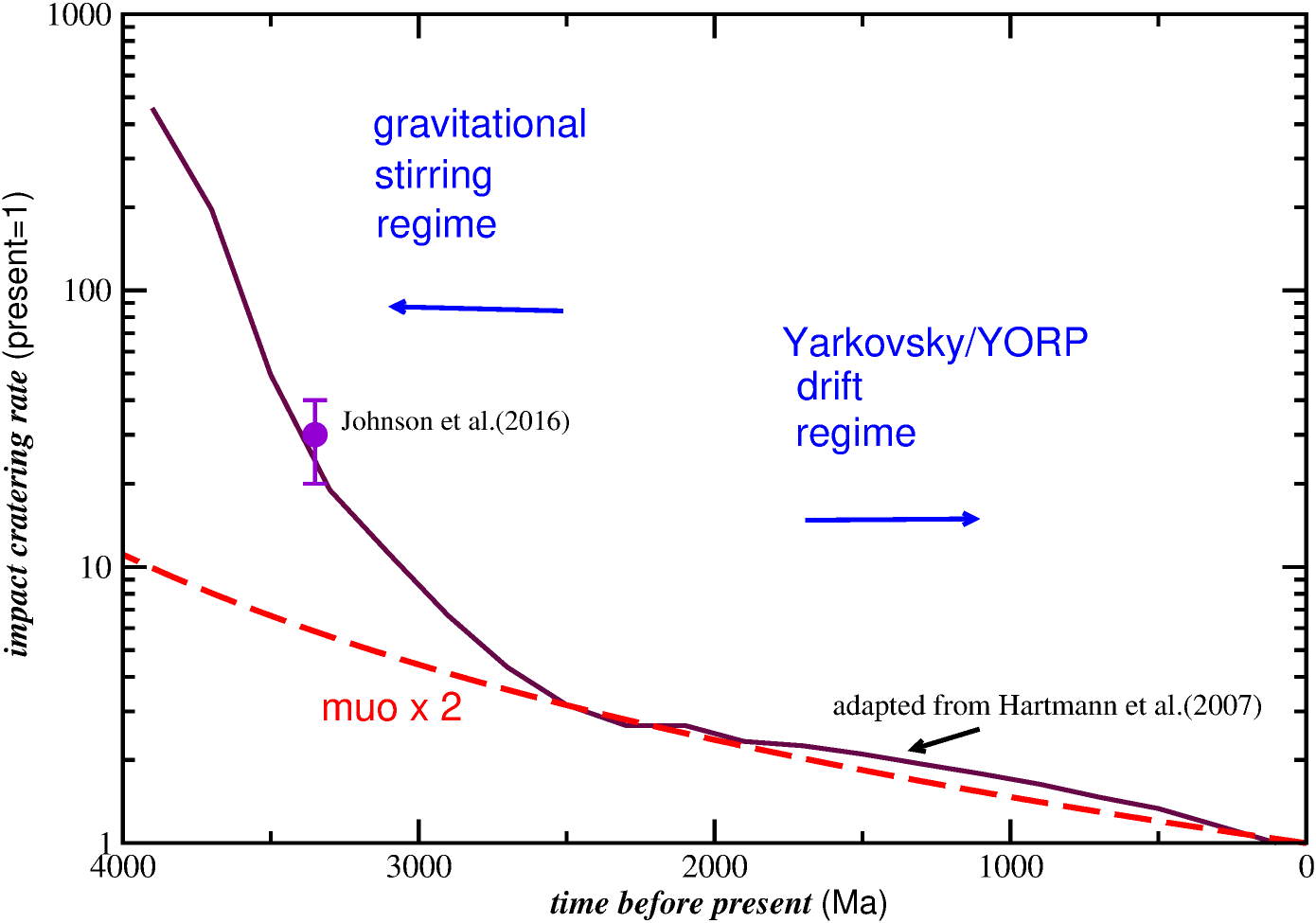}
\caption{Decline in the impact cratering rate of the Moon based on crater counts on Moon landing sites with well determined ages (solid curves). Results taken from \citet{Hart07}. The dashed curve is for our computed mass loss rate shown in Fig. \ref{belt_evol_prim} (lower panel) for 2$\mu_o$ normalized to one at $t=0$. We also plot an estimate of the impact rate of asteroids larger than 10 km with the Earth at about 3.25 Ga ago based on the discovery of strata of glass spherules of impact origin \citep{John16} (filled circle).}
\label{belt_evol_hart}
\end{figure}

How do the previous results match the observed impact cratering rate on the Moon and the Earth, assuming that they are proportional to the mass loss rate from the asteroid belt? As mentioned in the Introduction, in the first hundreds Ma after the solar system formed the asteroid belt may have been several orders of magnitude more massive than at present. From crater counts on the Moon at the Apollo/Luna missions sites, \citet{Hart07} found a fast decrease in the cratering rate down to about 3 - 2.5 Ga ago, followed by a smooth decrease afterwards (Fig. \ref{belt_evol_hart}). It is interesting to observe the good match between Hartmann et al's (2007) results and those obtained from layers of glass spherules, contained in the stratigraphy record, associated with several impacts of large asteroids ($D > 20$ km) that occurred around 3.5-3.2 Ga ago. The frequency of impacts of large asteroids at that time is estimated to have been between 30 and 50 times the current one \citep{Lowe10,John12,John16} (filled circle with error bars in Fig. \ref{belt_evol_hart}). We also have geologic evidence from Mars: from the study of the impact cratering rate on Martian landslides, \citet{Quan07} found that the cratering rate declined by a factor of three in the last 3 Ga. This result correlates very well with our computed mass loss rate for $2 \times \mu_o$ (Fig. \ref{belt_evol_prim}, bottom panel) at $\sim 2.5$ Ga ago.

Numerical simulations of the impact cratering rate of the Moon during the Archaean (4.0-2.5 Ga) by \citet{Nesv22} lead to a much flatter curve than that of Fig. \ref{belt_evol_hart} and a dominance of leftover planetesimals as the impactor source before 3.5 Ga. Yet, \citet{Marc21} find an under-recording of spherule beds in the late Archaean of up to a factor of ten, which suggests a much higher impact flux more in line with the results of \citet{Hart07}. Interestingly, \citet{Marc21} found a major shift in the surface chemistry of the Earth at $\sim 2.4-2.3$ Ga associated to the rise of oxygen in the atmosphere. Since impactors constitute an important sink of O$_2$, it is straightforward to relate the tail-end of the intense bombardment at $\sim 2.5$ Ga, shown in Fig. \ref{belt_evol_hart} with the rise of atmospheric O$_2$.

If we compare the slope of the impact cratering rate curve of Fig. \ref{belt_evol_hart} with that of the mass loss rate curve for $2 \times \mu_o$,  normalized to one at $t=0$, we find a good match up to $\sim 2.5$ Ga ago. This close correspondence suggests that our computed $\mu_o$ of eq.(14) ia a good approximation to the current rate of relative mass loss from the asteroid belt (a difference of only a factor of two). Going further to the past, the slope of the impact cratering rate curve changes abruptly and departs from our computed smooth curves derived from our simple theoretical model. The fast increase in the impact cratering rate obeys, we believe, to the increasing influence of gravitational interactions between bodies in the asteroid belt that increases the removal and injection of asteroids in the terrestrial planets zone. We will analyze next how this may happen.

\section{Removal of bodies from the asteroid belt: gravitational stirring versus Yarkovsky drift}

In order to be removed from the asteroid belt, bodies have to reach dynamical unstable niches like some MMRs with Jupiter (e.g. 3:1, 5:2, 2:1) or secular resonances like $\nu_6$, that pump up the eccentricities to unity \citep{Migl98}. There is an agreement that today the main responsible for moving fragments to any of these resonances is the Yarkovsky/YORP effect. 

Typical Yarkovsky drifts are $\sim 10^{-3} - 10^{-4}$ au Ma$^{-1}$ for a $D \sim 1$-km asteroid \citep{Broz06,Nuge12}. Since the Yarkovsky drift in semimajor axis goes as $\dot{a} \propto D^{-1}$, the larger the body, the less affected it will be. \citet{Carr17} were able to detect Yarkovsky drifts in members of the Veritas family up to (6374) Beslan ($D = 22.4$ km according to the JPL Solar System Dynamics) for which they found $|\dot{a}| \sim 10^{-4}$ au Ma$^{-1}$. Yet, they  were unable to find any appreciable Yarkovsky drift in (1086) Nata with an estimated $D \sim 66.3$ km. Objects greater than $D \sim 30$ km have negligible Yarkovsky drifts, so they will likely suffer a catastrophic collision before evolving to an unstable zone. We only have as potential escapees from the outer asteroid belt larger than 30 km the following objects: the Jupiter-crossers (944) Hidalgo ($D=38$ km), (6144) Kondojiro ($D=32.5$ km) and (20898) Fountainhills ($D=37.3$ km), though they could be as well interlopers from the Centaur population, Hildas or Trojans. The sizes of these potential escapees from the outer asteroid belt are consistent with those found for the largest NEAs and Mars-crossers (cf. Section 3.1).

In the early solar system, when the asteroid belt was much more massive than at present, gravitational stirring could have been much more important in moving large objects to resonances than the Yarkovsky mechanism. From numerical simulations of swarms of $\sim 10^{26}$-g planetary embryos, \citet{Weth92} found that gravitational scattering was the main mechanism for depleting the early asteroid belt. We can show that lunar-size bodies are not required to clear the asteroid belt. Smaller masses, say Pluto- or Ceres-size, can still produce considerable gravitational stirring in an early, more crowded asteroid belt. Let us consider a very simple model: an asteroid belt confined between heliocentric distances $2.1 < r < 3.3$ au containing asteroids with typical eccentricities $\sim 0.1$ and inclinations $\sim 10^{\circ}$ with a mass 10 times greater than at present. If the current number of asteroids with diameters $D > 1$ km in the main belt is $(1.2 \pm 0.5) \times 10^6$ \citep{Tede02}, then if we assume that the size distribution was similar in the past, a ten-fold asteroid belt would have contained about $1.2 \times 10^7$ asteroids with $D > 1$ km. The thickness of the asteroid belt is approximately $(r_1 + r_2)/2 \times \sin{i} \times 2$, where $r_1 = 2.1$ au, $r_2 = 3.3$ au, and $i=10^{\circ}$. Therefore, the volume $V$ of the asteroid belt is 

$$V \simeq \pi(r_2^2-r_1^2) \times (r_1 + r_2) \times \sin{i} = 6.44 \times 10^{25} \mbox{ km$^3$}. \mbox{\hspace{1cm} (20)}$$ 
The number density of asteroids is

$$n_{ast} = \frac{1.2 \times 10^7}{V} = 1.9 \times 10^{-19} \mbox{ km$^{-3}$}. \mbox{\hspace{1cm} (21)}$$  

The largest members of the asteroid belt will cause some dynamical stirring on neighbor small asteroids that can be described as impulses imparted by the largest members to the minor ones. The accumulation of small kicks will slightly change the orbits of the small bodies. Let us assume that the largest members of the early asteroid belt were -conservatively- Ceres-size of mass $M_C$. We can consider the two-body approximation in a close encounter of one object of mass $M_C$ with a small asteroid in which the latter will receive an impulse   

$$|\Delta v| = \frac{2GM_C}{ud_i}, \mbox{\hspace{1cm} (22)}$$  
where $G$ is the gravitational constant, $u$ the encounter velocity, and $d_i$ the impact parameter. If we assume that the massive asteroid moves on a circular orbit of radius $a_c$, while the interacting body has an orbit of semimajor axis $a$, eccentricity $e$ and inclination $i$, the encounter velocity will be given by \citep{Opik51}

$$U^2 = 3 - \frac{1}{A} - 2\sqrt{A(1-e^2)} \cos{i}, \mbox{\hspace{1cm} (23)}$$  
where $A=a/a_c$ is the semimajor axis of the small asteroid in units of the semimajor axis of $M_C$. We adopt $a=a_c$ so that $A=1$, $e=0.1$ and $i=10^{\circ}$, $U=u/v_{circ}$, where $v_{circ} = \sqrt{\mu/a_c}$, $\mu = GM_{\odot}$ and $M_{\odot}$ is the Sun's mass.

The orbital velocity $v$ of the small asteroid at the encounter with the massive asteroid is

$$v^2 = \mu\left(\frac{2}{r} - \frac{1}{a}\right) \simeq \frac{\mu}{a}. \mbox{\hspace{1cm} (24)}$$ 
Due to the impulse $\Delta v$, the small body will experience a change in its semimajor axis of the order

$$2v|\Delta v| = \frac{\mu}{a^2}|\Delta a| \mbox{\hspace{1cm} $\Longrightarrow$ } \hspace{1cm} |\Delta a| = 2\left(\frac{a^3}{\mu}\right)^{1/2} |\Delta v|. \mbox{\hspace{1cm} (25)}$$ 
Let us consider as an example the change $\Delta a$ caused by an encounter with an impact parameter $d_i = 10^4$ km. By introducing this value and $M_C = 9.4 \times 10^{23}$ g in eq.(22) and then solving eq.(25) for a typical $a=2.5$ au, we find $\Delta a \simeq 10^{-3}$ au. This is of the same order as the Yarkovsky drift found by \citet{Nuge12} for an asteroid of $D=1$ km.

Let us now estimate how many small asteroids are expected to be encountered within the target radius $d_i$ by the Ceres-size asteroid during a time span $\Delta t = 1$ Ma. We have

$$n_{enc} = \pi d_i^2 n_{ast} u \Delta t \mbox{\hspace{1cm} (26)}$$ 
and introducing the corresponding numerical values we obtain $n_{enc} = 7121$ bodies that will receive impulses able to cause drifts $\Delta a \gsim 10^{-3}$ au. We can compare our previous result derived from a simple analytical model with that obtained by \citet{Nesv02} from the numerical integration of a sample of 300 numbered asteroids that are assumed massless, except Ceres, Vesta and Pallas which are included with their actual masses. The dynamical evolution of the system is followed by 100 Ma considering the perturbation of the three massive asteroids and seven planets (from Venus to Neptune). They find that 50\% of the bodies drifted by more than $5 \times 10^{-4}$ au over 100 Ma. In our case, from eq.(26) we find that $\sim 7 \times 10^5$ asteroids will suffer encounters in 100 Ma that cause changes $\Delta a \gsim 10^{-3}$ au, namely $\sim 6$\% of the whole population ($1.2 \times 10^7$ objects). The number of encounters that are able to impart smaller changes $\Delta a = 5 \times 10^{-4}$ au will be three times greater, namely $\sim 20$\% of the asteroid population. The results are quite consistent with those of \citet{Nesv02} bearing in mind the different procedures and initial conditions, in particular in our case we adopted a somewhat more extended asteroid belt ($2.1 \lsim a(\mbox{au}) \lsim 3.3$ versus $2.1 \lsim a(\mbox{au}) \lsim 2.8$ for \citet{Nesv02}).

We can see that gravitational stirring by massive asteroids of mass $M_C$ affects a good number of small asteroids during a time span of 1 Ma. And this can be considered a lower limit because there could have been several Ceres-size or even larger asteroids in the early asteroid belt (say in the first Ga). The other great advantage of gravitational stirring, as compared to the Yarkovsky effect, is that the former is size-independent so while appreciable Yarkovsky drifts are limited to bodies with $D \lsim 30$ km, gravitational stirring does not have such an upper limit. Therefore, in the early solar system 100-km size bodies or even larger could have been displaced to unstable zones from where they were rapidly scattered. Once the population of leftover planetesimals subsided, asteroids took over as the main source of projecties (probably after $\sim 0.5-1$ Ga).

\section{Concluding remarks}

We can summarize our main results in the following points:

\begin{enumerate}

\item By assuming that 25\% of the zodiacal dust is of asteroidal origin, we find that the asteroid belt is losing a fraction $\mu_o \simeq 8.8 \times 10^{-5}$ Ma$^{-1}$ of its active mass (without Ceres, Vesta and Pallas).
\item We find that about 80\% of the mass that is lost from the asteroid belt leaves as meteoritic dust, and about 20\% as macroscopic bodies.
\item The extrapolation to the past of our canonical $\mu_o$ of eq.(14) would lead to a modest increase of the mass of the asteroid belt and the mass loss rate 3-3.5 Ga ago (about 50\% in the mass of the asteroid belt and a factor of two in the mass loss rate).
\item Given the uncertainties, if $\mu_o$ (or a time-average value over the last few hundreds of Ma) turns out to be two-three times greater than the value computed in eq.(14), its extrapolation to the past would lead to a dramatic increase, in both mass of the asteroid belt and the mass loss rate, 3-3.5 Ga ago, more in line with what is suggested by the geologic record of the Earth and the Moon.
\item In particular the slope of the mass loss rate curve of Fig. \ref{belt_evol_prim} for $2 \times \mu_o$ is found to fit very well with the slope of \citet{Hart07} impact cratering rate curve up to $\sim 2.5$ Ga in the past, For $t \gsim 2.5$ Ga both curves diverge suggesting that gravitational stirring in a more crowded asteroid belt took over as the dominant mechanism for moving bodies to unstable dynamical zones, thus favoring the scattering of asteroids in a  size-independent way that led to major collisional events.

\end{enumerate}  

\section{Appendix}

If the cumulative luminosity function, $N_H(<H)$, obeys a linear relation of the type

$$\log{[N_H(<H)]} = a +\alpha H \mbox{\hspace{1cm} (A1)}$$
and we assume that the luminosity $L$ of the asteroid is $L \propto D^2$, where $L$ is related to the magnitude $H$ by an equation of the type

$$H = c_o - 2.5\log{L} = c_1 - 2.5\log{D^2}, \mbox{\hspace{1cm} (A2)}$$
then we have

$$\log{[N_D(>D)]} = c' - 5\alpha\log{D}, \mbox{\hspace{1cm} (A3)}$$
where $c_o$, $c_1$ and $c'$ are constants.

The cumulative size distribution, $N_D(>D)$, will then follow a power-law of the form $N_D \propto D^{-s}$, where the exponent $s=5\alpha$. The differential size distribution will thus be given by

$$n_D = \frac{dN_D}{dD} = CD^{-(s+1)}, \mbox{\hspace{1cm} (A4)}$$
where $C$ is a constant that depends on the total population of asteroids $N_T$. If the population of asteroids follows the size distribution given by eq.(A4) in the size range $D_1 < D < D_2$ we have

$$N_T = \int^{D_2}_{D_1} n_D(D) dD = C\int^{D_2}_{D_1} D^{-(s+1)} dD,$$
that leads to

$$C = \frac{sN_T}{(D_1^{-s}-D_2^{-s})}. \mbox{\hspace{1cm} (A5)}$$ 

From the differential size distribution and bearing in mind that $D = (6/\pi\rho)^{(1/3)}M^{(1/3)}$, where $\rho$ is the bulk density of asteroids, we can obtain the differential mass distribution, $n_M(M)dM$, as follows

$$n_M(M)dM = n_D[D(M)]\frac{dD}{dM}dM$$
that leads to

$$n_M(M)dM = \frac{C}{3}\left(\frac{6}{\pi\rho}\right)^{-s/3} M^{-(s/3+1)} dM.  \mbox{\hspace{1cm} (A6)}$$
The total mass $M_T$ between $M_1<M<M_2$ will be given by,

$$M_T = \int_{M_1}^{M_2} Mn_M(M)dM =  \frac{C}{3}\left(\frac{\pi\rho}{6}\right)^{s/3} \int_{M_1}^{M_2} M^{-s/3}dM,$$
where $M_i = (\pi\rho/6)D_i^3$, $i=1,2$. By integrating this equation we obtain

$$M_T = \frac{C}{3}\left(\frac{\pi\rho}{6}\right)^{s/3} \left[\frac{M_2^{(1-s/3)}-M_1^{(1-s/3)}}{(1-s/3)}\right], \mbox{\hspace{1cm} (A7a)}$$
which is valid always that $s \neq 3$. For the case $s=3$ we get the alternative equation

$$M_T = \frac{C}{3} \left(\frac{\pi\rho}{6}\right)\ln{\left(\frac{M_2}{M_1}\right)} = C \left(\frac{\pi\rho}{6}\right) \ln{\left(\frac{D_2}{D_1}\right)}. \mbox{\hspace{1cm} (A7b)}$$

\bigskip

\centerline{\bf Acknowledgments}

\medskip

I want to thank the two anonymous referees for their helpful comments and suggestions.

\end{document}